\documentclass[aps,pra,onecolumn,superscriptaddress,groupedaddress]{revtex4}

\usepackage{amssymb}
\usepackage{color,graphicx}
\usepackage{amsmath}
\usepackage{amsbsy}
\usepackage{amsthm}
\usepackage{bbm}
\usepackage{bm}
\usepackage{epsfig}
\usepackage{lscape}
\usepackage{float}
\usepackage{subfigure}

\newcommand{\D}{\text{d}}

\begin{document}

\title{New avenues for testing collapse models}

\author{Andrea Vinante}
\email{a.vinante@soton.ac.uk}
\affiliation{School of Physics and Astronomy, University of Southampton, Southampton, SO17 1BJ, United Kingdom}

\author{Hendrik Ulbricht}
\email{h.ulbricht@soton.ac.uk}
\affiliation{School of Physics and Astronomy, University of Southampton, Southampton, SO17 1BJ, United Kingdom}

\date{\today}

\maketitle

\section{Introduction}

There is an increasing interest in developing experiments aimed at testing collapse models, in particular the Continuous Localization Model (CSL), the natural evolution of the GRW model initially proposed by Ghirardi et al \cite{GRW, CSL, collapse_review1, collapse_review2}. Current experiments and related bounds on collapse parameters are partially discussed in other contributions in this review. Our aim here is to discuss some of the most promising directions towards future improvements. The paper is organized as follows. In Sections~\ref{mechanical},~\ref{heating},~\ref{condensates} we will discuss noninterferometric techniques. In detail, in Section~\ref{mechanical} we will discuss mechanical experiments, both with conventional and levitated mechanical resonators, in Section~\ref{heating} we will consider proposed experiment looking at bulk thermal heating of solid bodies and in Section~\ref{condensates} we will briefly discuss the use of cold atoms or macroscopic condensates. In Section~\ref{matter} we will outline proposals of matter-wave interference with massive nano/microparticles. We will end in Section~\ref{other} with some ideas on how precision experiments can be used for testing collapse models.

\section{Noninterferometric mechanical tests of collapse models}   
\label{mechanical}

\subsection{Key concepts}

This class of experiments has emerged in recent years as one of the most powerful and effective ways to test collapse models. The underlying idea \cite{collett,adler2005,adler} is that a mechanism which continuously localizes the wavefunction of a mechanical system, which can be either a free mass or a mechanical resonator, must be accompanied by a random force noise acting on its center-of-mass. This leads in turn to a random diffusion which can be possibly detected by ultrasensitive mechanical experiments.

In a real mechanical system such diffusion will be masked by standard thermal diffusion arising from the coupling to the environment, i.e. from the same effects which lead to decoherence in quantum interference experiments \cite{arndt}. In practice there will be additional nonthermal effects, due to external nonequilibrium vibrational noise (seismic/acoustic/gravity gradient). Moreover, one has to ensure that the back-action from the measuring device is negligible.
 
Under the assumption that thermal noise is the only significant effect, the (one-sided) power spectral density of the force noise acting on the mechanical system is given by:
\begin{equation}\label{forcenoise}
  S_{ff}=\frac{4 k_B T m \omega}{Q} + 2 \hbar^2 \eta.
\end{equation}
where $k_B$ is the Boltzmann constant, $T$ is the temperature, $m$ is the mass, $\omega$ the angular frequency, $Q$ is the mechanical quality factor.

$\eta$ is a diffusion constant associated to spontaneous localization, and can be calculated explicitly for the most known models. For CSL, it is given by the following expression
\begin{align}\label{eta}
\eta &= \frac{2\lambda}{m_0^2}\,\iint\,\D^3{\bf r}\,\D^3\mathbf{r}'\,
\exp\left(-\frac{|\mathbf{r}-\mathbf{r}'|^2}{4r_C^2}\right)\,
\frac{\partial\varrho({\bf r})}{\partial{z}}\,\frac{\partial\varrho({\bf r}')}{\partial{z'}} \\&=
\frac{(4\pi)^{\frac32}\,\lambda\,r_C^3}{m_0^2}\,
\int\frac{\D^3{\bf k}}{(2\pi)^3}\,k^2_z\,e^{-{\bf k}^2r_C^2}\,|\tilde{\varrho}({\bf k})|^2
\end{align}
with ${\bf k}=(k_x,k_y,k_z)$, $\tilde{\varrho}({\bf k})=\int\D^3{\bf x}\,e^{i{\bf k}\cdot{\bf r}}\,\varrho({\bf r})$ and $\varrho({\bf r})$ the mass density distribution of the system. In the expressions above $m_0$ is the nucleon mass and $r_C$ and $\lambda$ are the free parameters of CSL. The typical values proposed in CSL literature are $r_C = 10^{-7}$ m and $10^{-6}$ m, while for $\lambda$ a wide range of possible values has been proposed, which spans from the GRW value $\lambda \approx 10^{-16}$ Hz \cite{GRW,CSL} to the Adler value $\lambda \approx 10^{-8 \pm 2}$ Hz at $r_C=10^{-7}$ m \cite{adler}. The possibility for such non-interferometric tests, which aim to directly test the non-thermal noise predicted by collapse models has been pointed out first by Bahrami et al.~\cite{Bahrami} and the ideas has been picked-up rapidly by many others~\cite{nimmrichter, Bera, diosi, goldwater, vinanteCSL1}.

An experiment looking for CSL-induced noise has to be designed in order to maximize the 'noise to noise' ratio between the CSL term and the thermal noise. In practice this means lowest possible temperature $T$, lowest possible damping time, or linewidth, $1/\tau = \omega/Q$, and highest possible $\eta/m$ ratio. The first two conditions express the requirement of lowest possible power exchange with the thermal bath, the third condition is inherently related to the details of the specific model. 

For CSL we can distinguish two relevant limits. When the characteristic size $L$ of the system is small, $L \ll r_C$, then the CSL field cannot resolve the internal structure of the system, and one finds $\eta/m \propto m$. When the characteristic length of the system in the direction of motion $L$ is large, $L \gg r_C$, then $\eta/m \propto \rho / L$, where $\rho$ is the mass density~\cite{nimmrichter, diosi, vinanteCSL1}. The expressions in the two limits imply that, for a well defined characteristic length $r_C$, the optimal system is a plate or disk with thickness $L \sim r_C$ and the largest possible density~$\rho$.

Among other models proposed in literature, we mention the gravitational Diosi-Penrose (DP) model, which leads to localization and diffusion similarly to CSL. The diffusion constant $\eta_{DP}$ is given by \cite{diosi}:
\begin{equation} \label{DP}
  \eta_{DP}=\frac{G \rho m}{6 \sqrt{\pi} \hbar} \left( \frac{a}{r_{DP}} \right)^3 , 
\end{equation}
where $a$ is the lattice constant and $G$ is the gravitational constant, so that he ratio $\eta_{DP}/m$ depends only on the mass density. Unlike CSL, there is no explicit dependence on the shape or size of the mechanical system.

\subsection{Cantilevers and other clamped resonators}

Experiments based on ultrasensitive cryogenic cantilevers have been historically the first serious attempt to bound collapse models using diffusive mechanical experiments. The micro/nanocantilever employed in these experiments are optimized devices developed in the context of atomic force microscopy. They are characterized by low stiffness, relatively low frequency $f_0 \sim $ kHz and high $Q$ factors in the range $10^5-10^7$. Operation at millikelvin temperature has been enabled by the use of SQUIDs for detection. 

Current cantilever-based experiments bound the CSL collapse rate to be lower than $10^{-8}$ Hz at $r_C=10^{-7}$ m, and $10^{-10}$ Hz at $r_C=10^{-6}$ m, values which are already partially excluding the Adler parameters \cite{vinanteCSL1, vinanteCSL2}.

It is not easy to push much further the current limits. Operation at lower temperature appears challenging due to increasing thermalization problems, while mechanical $Q$ can be hardly improved over current values due to clamping losses. It can be noticed that, unlike cantilevers, micromembranes with much higher quality factor up to $10^9$ have been demonstrated using an optimized design to suppress clamping losses \cite{schliesser}. However, these outstanding values are obtained only in high stress membrane at relatively high frequency $\sim 0.1-1$ MHz. The ratio $Q/f_0$ is not improved by this trick. Finally, pushing micromechanical systems to lower frequencies is possible but this approach has not been much investigated so far.

A novel route towards a significant improvement, specifically valid for the CSL model, has been recently proposed. The idea is to optimize the shape of a test mass to be attached on the cantilever, in order to maximize the effect at a given value of $r_C$ \cite{multilayer}. The proposed optimized shape is a multilayer structure, where many different layers of two alternate materials with large difference in mass density are stacked together. This configuration is predicted to enhance the effect of CSL for $r_C \lesssim 3 d$, where $d$ is the layer thickness, at the expense of reducing the effect at larger $r_C$. First experiments in this direction have been able to bound the CSL collapse rate well below $10^{-9}$ Hz at $r_C=10^{-7}$ m and are thus close to exclude completely the parameter range proposed by Adler \cite{multilayerexp}.

\subsection{Levitated particles}
\begin{figure}[!ht]
\includegraphics[width=\textwidth]{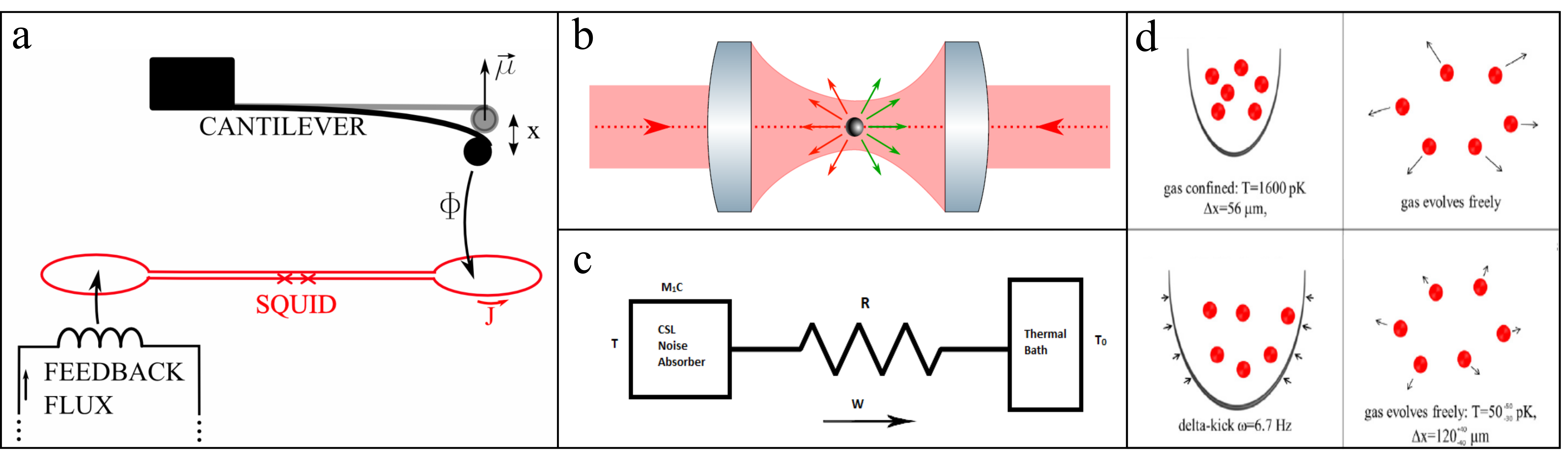}
\caption{Simplified sketch of some of the noninterferometric methods to test collapse models discussed in this contribution.
(a) Measuring the mechanical noise induced by CSL using an ultracold cantilever detected by a SQUID (adapted from Ref. \cite{vinanteCSL2}); (b) Measuring the mechanical noise induced by CSL using a levitated nanoparticle detected optically (adapted from Ref. \cite{TEQ}); (c) Measuring the heating induced by CSL in a solid matter object cooled to very low temperature (adapted from Ref. \cite{mishra}); (d) Measuring the increase of kinetic energy induced by CSL in a ultracold atoms cloud (adapted from Ref. \cite{bilardello1}).}   \label{scheme}
\end{figure}

One of the most promising approaches towards a significant leap forward in the achievable sensitivity to spontaneous collapse effects is by levitation of nanoparticles or microparticles. The main benefits of levitation are the absence of clamping mechanical losses and wider tunability of mechanical parameters. In addition, several degrees of freedom can be exploited, either translational or rotational \cite{goldwater, CSLrotational}. This comes at the price of higher complexity, poor dynamic range and large nonlinearities, which usually require active feedback stabilization over multiple degrees of freedom. However, levitated systems hold the promise of much better isolation from the environment, therefore higher quality factor. One relevant example, in the macroscopic domain, is the space mission LISA Pathfinder, which is based on an electrostatically levitated test mass, currently setting the strongest bound on collapse models over a wide parameter range \cite{lisa}.

Several levitation methods for micro/nanoparticles are currently being investigated. The most developed is optical levitation using force gradients induced by laser fields, the so called optical tweezer approach \cite{tweezer}. While this is a very effective and flexible approach to trap nanoparticles, in this context it is inherently limited by two factors: the relatively high trap frequency, in the order of $100$ kHz, and the high internal temperature of the particles, induced by laser power absorption, which leads ultimately to strong thermal decoherence. Alternative approaches have to be found, featuring lower trap frequency and low or possible null power dissipated in the levitated particle. The two possible classes of techniques are electrical levitation and magnetic levitation. 

Electrical levitation has been deeply developed in the context of ion traps. The standard tool is the Paul trap, which allows to trap an ion, or equivalently a charged nanoparticle, using a combination of ac and dc bias electric fields applied through a set of electrodes \cite{paultrap}. The power dissipation is much lower than in the optical case, and the technology is relatively well-established. However, the detection of a nanoparticle in a Paul trap still poses some technological challenge. 

This issue has been extensively investigated in a recent paper \cite{TEQ}, specifically considering a nanoparticle in a cryogenic Paul trap in the context of collapse model testing. Three detection schemes have been considered: an optical cavity, an optical tweezer, and a all-electric readout based on SQUID. It was found that to detect the nanoparticle motion with good sensitivity, optical detection has to be employed. Unfortunately, optical detection is not easily integrated in a cryogenic environment, and leads to a nonnegligible internal heating and excess force noise. On the other hand, an all-electrical readout would potentially allow for a better ultimate test of collapse models, but at the price of a very poor detection sensitivity, which could make the experiment hardly feasible. The authors argue that a Paul-trapped nanoparticle, with an oscillating frequency of 1 kHz, cooled in a cryostat at $300$ mK with an optical readout may be able to probe the CSL collapse rate down to $10^{-12}$ Hz at $r_C=10^{-7}$ m. A SQUID-based readout , if viable, could theoretically allow to reach $10^{-14}$ Hz.

A recent experiment employing a nanoparticle in a Paul trap with very low secular frequencies at $\sim 100$ Hz and low pressure has demonstrated ultranarrow linewidth $\gamma/2\pi=82$~$\mu$Hz \cite{pontin}. This result has been used to set new bounds on the dissipative extension of CSL. This experiment may be able to probe the current limits on the CSL model in the near future, once it will be performed at cryogenic temperature and the main sources of excess noise, in particular bias voltage noise, will be removed.

Magnetic levitation, while less developed, has the crucial advantage of being completely passive. Furthermore the trap frequencies can be quite low, in the Hz range. Three possible schemes can be devised: levitation of a diamagnetic insulating nanoparticle with strong external field gradients \cite{durso, cinesi}, levitation of a superconducting particle using external currents \cite{oriol1, leadzeppelin}, and levitation of a ferromagnetic particle above a superconductor \cite{oriol2}. 

The first approach has been recently considered in the context of collapse models \cite{cinesi}. The experiment was based on a polyethylene glycol microparticle levitated in the static field generated by neodymium magnets and optical detection. The experiment has been able to set an upper bound on the CSL collapse rate $\lambda<10^{-6.2}$ Hz at $r_C=10^{-7}$ m, despite being performed at room temperature. A cryogenic version of this experiment should be able to approach the current experimental limits on CSL.

The second and third approach based on levitating superconducting or ferromagnetic particles are currently investigated by a handful of groups \cite{oriol1, leadzeppelin, oriol2, chris, me}, but no experiment has so far reached the experimental requirements needed to probe collapse models. However, a significant progress has been recently achieved: a ferromagnetic microparticle levitated above a type I superconductor (lead) and detected using a SQUID, has demonstrated mechanical quality factors for the rotational and translational rigid body mechanical modes exceeding $10^7$, corresponding to a ringdown time larger than $10^4$ seconds \cite{me}. The noise is this experiment is still dominated by external vibrations. However, as the levitation is completely passive and therefore compatible with cryogenic temperatures, this appears as an excellent candidate towards near future improved tests of collapse models.

\section{Bulk heating experiments}
\label{heating}

A different strategy towards testing the violation of energy conservation caused by collapse models is to search for anomalous heating of specific systems. The difference compared to mechanical experiments is that in the latter case one looks for an energy increase in an individual degree of freedom, which is a purely mechanical effect, while in the general case one looks for the total increase of internal energy (i.e. of temperature) of a macroscopic body, which is a thermal effect. 

According to Adler \cite{adler}, the heating of a macroscopic body   due to CSL can be generally written as:
\begin{equation}
 \frac{dE}{dt}=\frac{3}{4}\frac{\lambda \hbar^2 M}{r_C^2 m_N^2}
\end{equation}
where $M$ is the total mass of the system.
This expression, initially derived for a gas of noninteracting particles, has been shown to be very general. For instance, it holds for standard solid state systems \cite{adlervinante} and for   nonstandard matter such as Fermi liquids \cite{neutron}. 

An important caveat has to be pointed out in the case of condensed matter systems, in particular of solids. The $\lambda$ factor has to be regarded as an effective value $\lambda_{\mathrm{eff}}$, averaged over the frequencies of the internal (phononic) modes of the system. This average value is shown to be $\lambda_{\mathrm{eff}} \approx \lambda \left( \omega_0 \right)$ , where $\omega_0$ is the frequency corresponding to phonons with wavelength $\sim r_C$ \cite{adlervinante, bahrami}. For  typical solid matter, this corresponds to $\omega_0 \approx 10^{11}$~$s^{-1}$. A consequence of this fact is that any bound from heating of solid matter would be evaded by a nonwhite CSL noise with a low-pass cutoff at frequency lower than $\omega_0$.

As the bulk heating scales with the mass, one possible experimental approach is to estimate this effect in astronomical objects. For instance by analyzing the intergalactic medium, mainly composed of cold hydrogen, one can infer a bound on the CSL collapse rate $\lambda < 10^{-8}$~Hz \cite{adler} at $r_C=10^{-7}$~m (in the following of this chapter we will always assume $r_C=10^{-7}$~m). More recently, it has been suggested that neutron stars can set much stronger bounds \cite{tilloy}. The actual bounds inferred from current observational data are however not yet competitive, at level $\lambda_{\mathrm{eff}}<10^{-7}$~Hz \cite{neutron, tilloy}, but speculative bounds based on the capabilities of future astronomical surveys suggest that much stronger bounds can be obtained in the future. 
Stronger bounds, at level $\lambda_{\mathrm{eff}}<10^{-10}$~Hz, can be inferred from the astronomical data on planets of the solar system, in particular Neptune \cite{neutron}. This is essentially due to the very low temperature of these planets. An even better bound, at level $\lambda_{\mathrm{eff}}<10^{-11}$~Hz, can be inferred from the earth thermal balance, once primordial and radiogenic sources of Earth heat are very carefully taken into account~\cite{adlervinante}. A more speculative prospect is to test collapse models by evaluating their effect at cosmological level, for instance in the Cosmic Microwave Background. 

Here, we focus instead on the possibility of detecting very small heating in controlled laboratory experiments. Bulk massive objects can be routinely cooled down to very low temperatures. Dilution refrigerators can be used to cooldown relatively massive objects down to $\sim 10$ mK. The most massive object ever cooled in this way is probably the CUORE detector looking at neutrinoless beta decay \cite{CUORE}, with a mass of $\sim1$ ton cooled to $10$ mK. Much lower temperatures, even below $100$ $\mu$K, can be reached by adiabatic nuclear demagnetization cryostats \cite{pobell}. Here, the typical mass which can be cooled is of the order of several kilograms.

As the thermalization of any object becomes increasingly difficult at lower and lower temperature, a crucial requirement of these experiments is to suppress as much as possible any heat leak. The dominant residual heat sources are (i) vibrations, (ii) relaxation of internal stress and two-level systems and (iii) the background of radioactivity and muons from cosmic rays \cite{pobell}. The first two sources can be efficiently suppressed by proper mechanical isolation, proper choice of materials and by waiting for long relaxation times. Overall, the best residual heat leak  estimated in current experiments is of the order of $10^{-11}$ W/kg, limited by background muons, which corresponds to $\lambda_{\mathrm{eff}}<3 \times 10^{-11}$~Hz \cite{adlervinante}.

It is important to note that, due to the high penetration of cosmic muons, their background heating scales with the experimental mass and is therefore a fundamental barrier, unless the experiment is performed heavily underground. This idea has been considered recently by Mishra et al \cite{mishra}, who have estimated the achievable upper limit on $\lambda_{\mathrm{eff}}$ which could be detected by an ideal cosmic-background-limited detector placed underground. It has been found that the shielding provided by the deepest existing underground laboratory (the China Jinping Underground Laboratory, placed at 6.7 km of "water-equivalent" depth) would be sufficient to test the CSL model down to the GRW parameters, i.e $\lambda_{\mathrm{eff}} \approx 10^{-16}$~Hz. Slightly worse performance is expected by operating in alternative sites, such as the Gran Sasso Laboratory in Italy.

Of course such an experiment would require a systematic and very efficient suppression of any parasitic heating source, such as vibrations or internal relaxation, by several orders of magnitude. This appears a tough challenge, which is however purely technical and not related to fundamental limits.
The technology developed for existing underground cryogenic experiments looking for Dark Matter or neutrinoless double beta decay, such as CUORE \cite{CUORE}, is probably already good enough to push the current bounds by 1-2 orders of magnitude. However, a specific experimental design is needed, specifically optimized to detect a dc heating source such as CSL. In particular, thermometry with very good absolute accuracy is needed, in contrast with existing detectors which are optimized for very high responsivity in order to resolve individual high energy events.

\section{Cold atoms and condensates}
\label{condensates}

Cold atoms represent another possible system to detect effects related with collapse models, due to (i) the very low kinetic temperatures that can be achieved, which can be as low as a few pK and (ii) the flexibility of these systems which allow the realization of a variety of quantum states involving a relatively large number of atoms, in particular Bose-Einstein Condensates (BEC). An obvious drawback, which partially compensates these advantages, is the much lower density, i.e. a comparatively lower total number of atoms compared to solid state systems.

Some first investigations of using cold atoms in order to probe collapse models, in particular CSL, have been reported in literature \cite{pearleBEC, kasevich, bilardello1} and include an outlook on future improvements.

The first approach considered in Ref.~\cite{pearleBEC} consists in analyzing the lifetime of BEC condensates. The spontaneous heating induced by CSL results into an exponential decay of the ground state population, which can be bounded by experiments. Analysis of current experiments yielded a bound on the CSL collapse rate $\lambda < 10^{-7}$~Hz at the standard $r_C=10^{-7}$~m. The authors remarks that the analysis is provisional, and that new experiments specifically tailored to estimate the rate of energy increase will significantly improve the bounds. Specific improvements will be the use of heavy atomic mass (like cesium), lower background of foreign atoms, suppression of three-body recombination, and a sufficiently high barrier to eliminate evaporative cooling.

A second approach was investigated by Bilardello et al~\cite{bilardello1}, by analyzing an experiment performed by Kasevich et al~\cite{kasevich} with a diluted cloud of ultracold rubidium atoms. The last stage of this experiment consisted in a delta-kick optical-lensing cooling, which enabled free evolution of the cloud on a time scale of seconds at an extremely low temperature below $100$ pK. This bounds the CSL collapse rate to $\lambda<5 \times 10^{-8}$ Hz at $r_C=10^{-7}$ m, slightly improving BEC limits. The authors of the experiment estimate that this technique can be improved by 2-3 orders of magnitude before reaching fundamental limits imposed by diffraction limited collimation temperature \cite{kasevich}. The analysis performed in \cite{bilardello1} also shows that cold atoms experiments are particularly efficient in testing non Markovian extensions of the CSL model, such as the dissipative CSL model.

A third option is to use cold atoms to perform interferometric experiments. This will be discussed in the next section.

\section{Matter-wave Interferometry} \label{matter}

\begin{figure}[!ht]
\includegraphics[width=\textwidth]{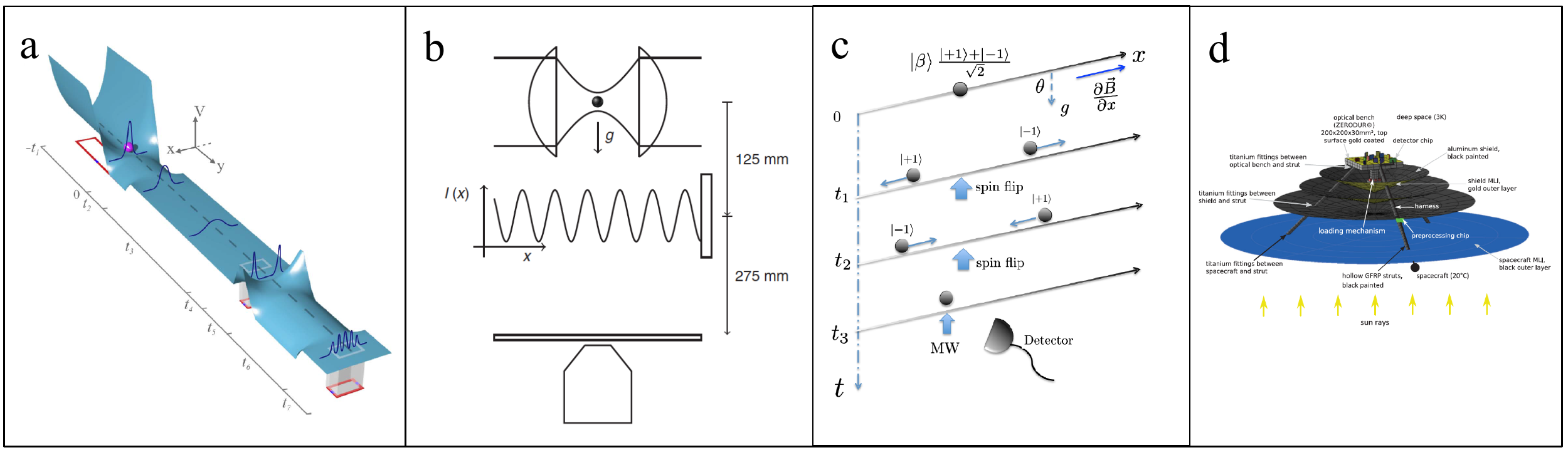}
\caption{Illustration of some of the proposed schemes for matterwave interferometry with nano- and micro-particles to test the quantum superposition principle directly, and therefore also collapse models.
(a) The cryogenic skatepark for a single superconducting micro-particle (adapted from Ref.~\cite{skatepark}); (b) The nanoparticle Talbot interferometer (adapted from Ref.~\cite{Bateman}); (c) The Ramsey scheme addressing the electron Spin of a NV-centre diamond coupled to an external magnetic field gradient $(\partial B/\partial x)$  (adapted from Ref.~\cite{Bose}); (d) The adaptation of an interferometer at a free falling satellite platform in space to allow form longer free evolution times (adapted from Ref.~\cite{maqro}).}   \label{interscheme}
\end{figure}

Matterwave interferometry is directly testing the quantum superposition principle. Relevant for mass-scaling collapse models, such as CSL, are matterwave interferometers testing the maximal macroscopic extend in terms of mass, size and time of spatial superpositions of single large-mass particles. Such beautiful, but highly challenging experiments have been pushed by Markus Arndt's group in Vienna to impressive particle masses of $10^4$ atomic mass units (amu), which is still not significantly challenging CSL. More details can be found in chapter {\bf XXX} of this volume. Therefore the motivation remains to push matterwave interferometers to more macroscopic systems. Predicted bounds on collapse models set by large-mass matterwave interferometers are worked out in detail in~\cite{Marko}.

As usual in open quantum system dynamics treatments, non-linear stochastic extensions of the Schr\"odinger equation on the level of the wavefunction~\cite{Zoller} correspond to a non-uniquely defined master equation on the level of the density matrix $\rho$ to describe the time evolution of the quantum system, say the spatial superposition across distance $|x-y|$, where the conserving von Neumann term $\partial\rho_t(x, y)/\partial t=-(i/\hbar) [H, \rho]$, is now extended by a Lindblad operator $L$ term:
\begin{equation} 
\frac{\partial\rho_t(x, y)}{\partial t} = -\frac{i}{\hbar}[H, \rho_t(x, y)] + L\rho_t(x, y), 
\end{equation}
where $H$ is the Hamilton operator of the quantum system and different realisations of a Lindblad operator are used to describe both standard decoherence (triggered by the immediate environment of the quantum system)~\cite{Breuer} as well as spontaneous collapse of the wavefunction triggered by the universal classical noise field as predicted by collapse models. 

Now the dynamics of the system is very different with and without the Lindbladian, where with the Lindbladian the unitary evolution breaks down and the system dynamics undergoes a quantum-to-classical transition witnessed by a vanishing of the fringe visibility of the matterwave interferometer. In the state represented by the density matrix the off-diagonal terms vanish as the system evolves according to the open system dynamics, the coherence/superposition of that state is lost. The principal goal of interference experiments with massive particles is then to explore and quantify the relevance of the $(L\rho_t(x, y))$-term - as collapse models predict a break down of the quantum superposition principle for a sufficient macroscopic system. An intrinsic problem is the competition with known and unknown environmental decoherence mechanisms, if a visibility loss is observed. However solutions seem possible.

In order to further increase the macroscopic limits in interference some ambitious proposals have been made utilizing nano- and micro-particles, c.f. Fig.(\ref{interscheme}). The main challenge is to allow for a long enough free evolution time of the prepared quantum superposition state in order to be sensitive to the collapsing effects. The free evolution - the spatial spreading of the wavefunction $\Psi(r, t)$ with time - according to the time-dependent Schr\"odinger equation with the potential $V(r)=0$,
\begin{equation}
\frac{\partial}{\partial t}\Psi(r, t) = -i\frac{\hbar}{2m} \nabla^2 \Psi(r, t),
\end{equation}
describes a diffusive process for probability amplitudes similar to a typical diffusion equation with the imaginary diffusion coefficient ($-i\hbar/2m$). Therefore the spreading of $\Psi(r, t)$ scales inversely with particle mass $m$. For instance for a $10^7$ amu particle it already takes so long to show the interference pattern in a matterwave experiment that the particle would significantly drop in Earth's gravitational field, in fact it would drop on the order of 100 m. This requires a dramatic change in the way large-mass matterwave interferometry experiments have to be performed beyond the mass of $10^6$ amu~\cite{Bateman}.

Different solutions are thinkable. One could of course envisage building a 100 m fountain, but that seems very unfeasible also given that no sufficient particle beam preparation techniques exist (and don't seem to be likely to be developed in the foreseeable future) to enable the launch and detection of particles in the mass range in question over a distance of 100 m. One can consider to levitate the particle by a force field to compensate for the drop in gravity, but here we face a high demand on the fluctuations of that levitating field, which have to be small compared to the amplitudes of the quantum evolution. This requirement does not appear to be feasible with current technology. A maybe possible option is to coherently boost/accelerate the evolution of the wavefunction spread by a beam-splitter operation. The proposals in Refs.~\cite{skatepark, Bose} are such solutions, which are still awaiting their technical realisation for large masses. A more realistic alternative, given current technical capabilities, is to allow for long enough free evolution by freely fall the whole interferometer apparatus in a co-moving reference frame with the particle. This is the idea of the MAQRO proposal, a dedicated satellite mission in space to perform large-mass matterwave interference experiments with micro- and nano-particles~\cite{maqro}.

Another interesting approach is to consider the use of cold or ultra-cold ensembles of atoms such as cloud in a magneto optical trap (MOT) or an atomic Bose-Einstein Condensate (BEC) as also there we find up to $10^8$ atoms of alkali species such as rubidium or caesium. On closer look it turns out that such weakly interacting atomic ensembles are not of immediate use for the purpose to test macroscopic quantum superpositions in the context of collapse model test. For instance, a crucial property for testing the CSL model is the mass-proportional (number of particles $N$, more precisely the number of nucleons: protons and neutrons in the nuclei of the atoms) amplification which in principle can even go with $N^2$. This effect can be seen in Eq.~(\ref{eta}), and can be thought as arising from the classical collapse noise (treated as a wave with correlation length $r_c$) coherently scattering off the particle in the quantum superposition state. Naively, if the CSL noise is collapsing the wavefunction of only one of the constituent nucleons, then the total wavefunction of the whole composite object collapses. While this holds for a solid nanoparticle consisting of many atoms (and therefore nucleons), it is not the case for a weakly interacting atomic ensemble. If one atom is collapsing then the total atomic wavefunction remains intact and the one atom is lost from the ensemble.

This may change if the atoms in the cold or ultra-cold ensemble can be made strongly interacting, without running into the complications of chemistry which may forbid condensation of the atomic - then molecular - cloud at all. However there is hope to circumvent this problem by means of quantum optical state preparation techniques applied after a BEC has been formed. For instance, collective NOON or squeezed states, featuring macroscopic entanglement between individual atoms, would enable $N$ and even $N^2$ scaling in the fashion fit for testing wavefunction collapse. This approach is extremely challenging, and is discussed in detail in Ref.~\cite{bilardello2}.

A different scenario might arise if the physical mechanism responsible for the collapse of the wavefunction, which remains highly speculative at present, is in any way related to gravity~\cite{probe}, then there might be hope that atomic ensembles even in the weakly interacting case can be used to test CSL-type models. The condition to fulfil is that the atomic ensemble is interacting gravitationally strong enough so that it acts collectively under collapse, even if just a single constituent atom (nucleon) is affected by the collapsing effect. That hope is possibly very weak.

\section{Some concluding remarks} \label{other}
We have discussed avenues for non-interferometric and interferometric tests of the linear superposition principle of quantum mechanics in direct comparison to predictions from collapse models which break the linear/unitary evolution of the wavefunction. As matters stand both non-interferometric and interferometric set already bounds on the CSL collapse model, while those from non-interferometric tests are stronger by orders of magnitude. The simple reason lies in the immense difficulty to experimentally generate macroscopic superposition states, however a number of proposals have been made and experimentalists are set to approach the challenge.

We want to close by mentioning that there are possibly other experimental platforms which could set experimental bounds on collapse models and it would be of interest to study those in detail. Collapse models predict a universal classical noise field to fill the Universe and in principle couple to any physical system. In the simplest approach the experimental test particle can be regarded as a two-level system, as typically described in quantum optics. Then the collapse noise perturbs the two-level system and emissive broadening and spectral shifts can be expected, unfortunately out of experimental reach at the moment~\cite{two}. The minuscule collapse effect on a single particle (nucleon) needs some sort of amplification mechanism which usually comes with an increase of the number of constituent particles. 
However, ultra-high precision experiments have improved a lot in recent years. For instance much improved ultra-stable Penning ion traps are used to measure the mass of single nuclear particles, such as the electron, proton, and neutron, with an ultra-high precision to test quantum electrodynamics predictions~\cite{Blaum}. In principle also here the effect of collapse models should become apparent. Any theoretical predictions are difficult as relativistic versions of collapse models still represent a serious formal challenge. Other high potentials for testing collapse are ever more precise spectroscopies of simple atomic species with analytic solutions such as transitions in hydrogen~\cite{Haensch} and needless to say atomic clocks~\cite{Ye}.
 
As tests move on to set stronger and stronger bounds, we have to remain open to actually find something new. It is so easy to disregard tiny observed effects as unknown technical noise. In the case of direct testing collapse noise it is a formidable theoretical challenge to think about possible physics responsible for collapse, satisfying the constrains given by the structure of the collapse equation: the noise has to be classical and stochastic. Such concrete physics models will predict a clear frequency fingerprint, should we ever observe the collapse noise field.

\section*{Acknowledgements}
We acknowledge support from the EU project TEQ (grant agreement 766900). We would like to thank our collaborators, discussion and debating partners on the topic of experimental testing collapse models for many enlightening discussions over the years: Angelo Bassi, Steve Adler, Mauro Paternostro, TP Singh, Marko Toro$\mathrm{\check{s}}$, Matteo Carlesso, Giulio Gasbarri, Sandro Donadi, Luca Ferialdi, Marco Bilardello, Catalina Curceanu, Andrew Briggs, Caslav Bruckner, Markus Arndt, Daniel Bedingham, Ward Stuyve, Kinjalk Lochan, Michael Drewsen, Peter F. Barker, Antonio Pontin, Anis Rahman, James Bateman, Sougato Bose, Myungshik Kim, Mohammad Bahrami, Stefan Nimmrichter, Klaus Hornberger, Andrew Steane, Oriol Romero-Isart and Tjerk Oosterkamp. The paper is dedicated to the late Gian Carlo Ghirardi, who had a great idea and fought for it all his life.

\end{document}